\documentclass[preprint]{aastex}
\usepackage{float}
\usepackage{color}

\title{Interstellar Carbodiimide (HNCNH) - A New Astronomical Detection from the GBT PRIMOS Survey via Maser Emission Features}
\author{Brett A. McGuire}
\affil{Division of Chemistry and Chemical Engineering, California Institute of Technology, Pasadena, CA 91125}
\author{Ryan A. Loomis}
\affil{Department of Chemistry, University of Virginia, Charlottesville, VA, 22904}
\author{Cameron M. Charness \& Joanna F. Corby}
\affil{Department of Astronomy, University of Virginia, Charlottesville, VA, 22904}
\author{Geoffrey A. Blake}
\affil{Division of Chemistry and Chemical Engineering and Division of Geological and Planetary Sciences, California Institute of Technology, Pasadena, CA 91125}
\author{Jan M. Hollis}
\affil{NASA Goddard Space Flight Center, Greenbelt, MD, 20771}
\author{Frank J. Lovas}
\affil{National Institute of Standards and Technology, Gaithersburg, MD 20899}
\author{Philip R. Jewell \& Anthony J. Remijan}
\affil{National Radio Astronomy Observatory, Charlottesville, VA 22903}

\begin{document}

\begin{abstract}

In this work, we identify carbodiimide (HNCNH), which is an isomer of the well-known interstellar species cyanamide (NH$_2$CN), in weak maser emission, using data from the GBT PRIMOS survey toward Sgr B2(N).  All spectral lines observed are in emission and have energy levels in excess of 170 K, indicating that the molecule likely resides in relatively hot gas that characterizes the denser regions of this star forming region. The anticipated abundance of this molecule from ice mantle experiments is $\sim$10\% of the abundance of NH$_2$CN, which in Sgr B2(N) corresponds to $\sim$ 2 x 10$^{13}$ cm$^{-2}$.  Such an abundance results in transition intensities well below the detection limit of any current astronomical facility and, as such, HNCNH could only be detected by those transitions which are amplified by masing.   

\end{abstract}

\section{Introduction}

Historically, searches for new astronomical molecules resulted in the detection of favorable, high line strength transitions based on a thermal approximation to the excitation of these species in interstellar environments.  At the temperatures of hot molecular cores inside molecular clouds, these transitions often reside at (sub)millimeter wavelengths.  However, line confusion near the Boltzmann peak can lead to ambiguous identifications,  and the peak intensities of low abundance, large organic molecules may never rise above the noise floor or the confusion limit.  The success of molecule searches at centimeter wavelengths has shown that unique excitation conditions can lead to the unambiguous identification of very low abundance species with high accuracy. 

Carbodiimide (HNCNH) is the second most stable isomer of cyanamide (NH$_2$CN), with cyanamide being more stable by $\sim$4.0 kcal mol$^{-1}$ \citep{Duvernay2004,Tordini2003}.  Since the detection of NH$_2$CN by Turner et al. (1975) toward the high-mass star-forming region Sgr B2(N), HNCNH has been recognized as a candidate interstellar molecule.  HNCNH is also a tautomer of NH$_2$CN, and can be formed via an isomerization reaction whereby a H atom can migrate along the molecular backbone from the amine group.  At room temperature, HNCNH is in thermally-induced equilibrium with NH$_2$CN at a fraction of $\sim$1\%.  Given typical abundances of NH$_2$CN and temperatures in many astrophysical environments of $<< 300$ K, gas-phase tautomerization of NH$_2$CN is unlikely to produce HNCNH in appreciable abundance.  Yet, HNCNH may exist in detectable abundance out of thermal equilibrium with NH$_2$CN as in the case of interstellar hydrogen cyanide (HCN) and interstellar hydrogen isocyanide (HNC) - in which HNC is in greater abundance in some clouds (see Allen, Goddard, \& Schaefer 1980 and references therein).  

Although tautomerization is likely inefficient in the gas phase under interstellar conditions, experimental studies have shown NH$_2$CN $\rightarrow$ HNCNH conversion in water ices and matrices is far more efficient (Duvernay et al. 2005 and references therein). In the solid phase, the association of up to five water molecules with NH$_2$CN on an ice surface has been shown to significantly lower the activation barrier to tautomerization and promote the production of HNCNH \citep{Duvernay2005}.  In fact, HNCNH formation on a water-ice surface is shown to occur at temperatures as low as 70 K.  The relative abundance of HNCNH formed from this process has been measured to range from 4\% of NH$_2$CN at 80 K to as much as 13\% at 140 K \citep{Duvernay2005}.  In the laboratory, HNCNH sublimation occurs between 80 - 170 K, with no additional desorption observed above 170 K \citep{Duvernay2004}.  In the ISM, non-thermal desorption via shocks is also a likely liberation mechanism.  Assuming that tautomerization of NH$_2$CN on dust grain ice mantles is the dominant formation pathway for HNCNH, and that subsequent desorption occurs for both species, the gas phase abundance of HNCNH may be capped at $\sim$$10\%$ of NH$_2$CN.

NH$_2$CN has an estimated column density of $\sim$2$\times$10$^{14}$ cm$^{-2}$ towards Sgr B2(N) \citep{Nummelin2000}; we therefore anticipate an HNCNH column density on the order of ($\sim$10$^{13}$ cm$^{-2}$).  If thermal emission at hot core temperatures dominates the spectrum of HNCNH, the most favorable high line strength transitions are at millimeter and submillimeter wavelengths.  Nevertheless, we have previously used the Robert C. Byrd Green Bank Telescope (GBT) to detect new molecular species including trans-methyl formate \citep{Neill2012}, and cyanoformaldehyde \citep{Remijan2008} which \textit{only} had measurable astronomical line intensities detected at centimeter wavelengths.  Encouraged by these results, we searched for centimeter wave transitions of HNCNH.  All data were taken as part of the PRebiotic Interstellar MOlecular Survey (PRIMOS), and targeted rotation-torsion transitions of HNCNH selected from the Cologne Database for Molecular Spectroscopy\footnote{Original laboratory data cataloged from Birk et al. (1980), Wagener et al. (1995), \& Jabs et al. (1997).}$^,$\footnote{Available at www.splatalogue.net \citep{Remijan2008}.} \citep{Muller2005}.  Our results are presented in $\S$2.  In $\S$3, we evaluate the probability of mis-assigning these features to HNCNH and discuss evidence that observed transitions are masing.  Further, we explore the usefulness of our technique as a method for new molecule detection.

\section{Observations and Results}
\label{obs}

All data used for this project were taken as part of the National Radio Astronomy Observatory's (NRAO) 100-m Robert C. Byrd Green Bank Telescope (GBT) PRebiotic Interstellar MOlecule Survey (PRIMOS) Legacy Project.\footnote{Access to the entire PRIMOS dataset and specifics on the observing strategy including the overall frequency coverage, is available at http://www.cv.nrao.edu/$\sim$aremijan/PRIMOS/.}  This NRAO key project started in Jan 2008 and concluded in July 2011.  The PRIMOS project recorded a nearly frequency continuous astronomical spectrum from 1 to 50 GHz towards the Sgr B2(N) molecular cloud, with the pointing position centered on the Large Molecule Heimat (LMH) at (J2000) $\alpha$ = 17h47m19.8s, $\delta$ = -28$^{\circ}$22'17''.  An LSR source velocity of +64 km s$^{-1}$ was assumed.  Intensities are presented on the T$_A^*$ scale \citep{Ulich1976}.  See Neill et al. (2012) for further details of the observations, full data reduction procedure, and analysis.

\begin{figure}
\caption{Carbodiimide (HNCNH) spectral passbands toward Sgr B2(N) recorded from the GBT PRIMOS Survey. Rotation-torsion doublet transition quantum numbers are shown in each panel. The passband width displayed is 500 km s$^{-1}$ in each case.  The spectra are plotted as a function of frequency (MHz), corrected for a LSR source velocity of +64 km s$^{-1}$.  The blue and red vertical lines indicate the location of the transition rest frequency (see Table 1) at an assumed LSR source velocity of +64 km s$^{-1}$ and +82 km s$^{-1}$, respectively.  Data in all panels were Hanning smoothed for display purposes.}
\plotone{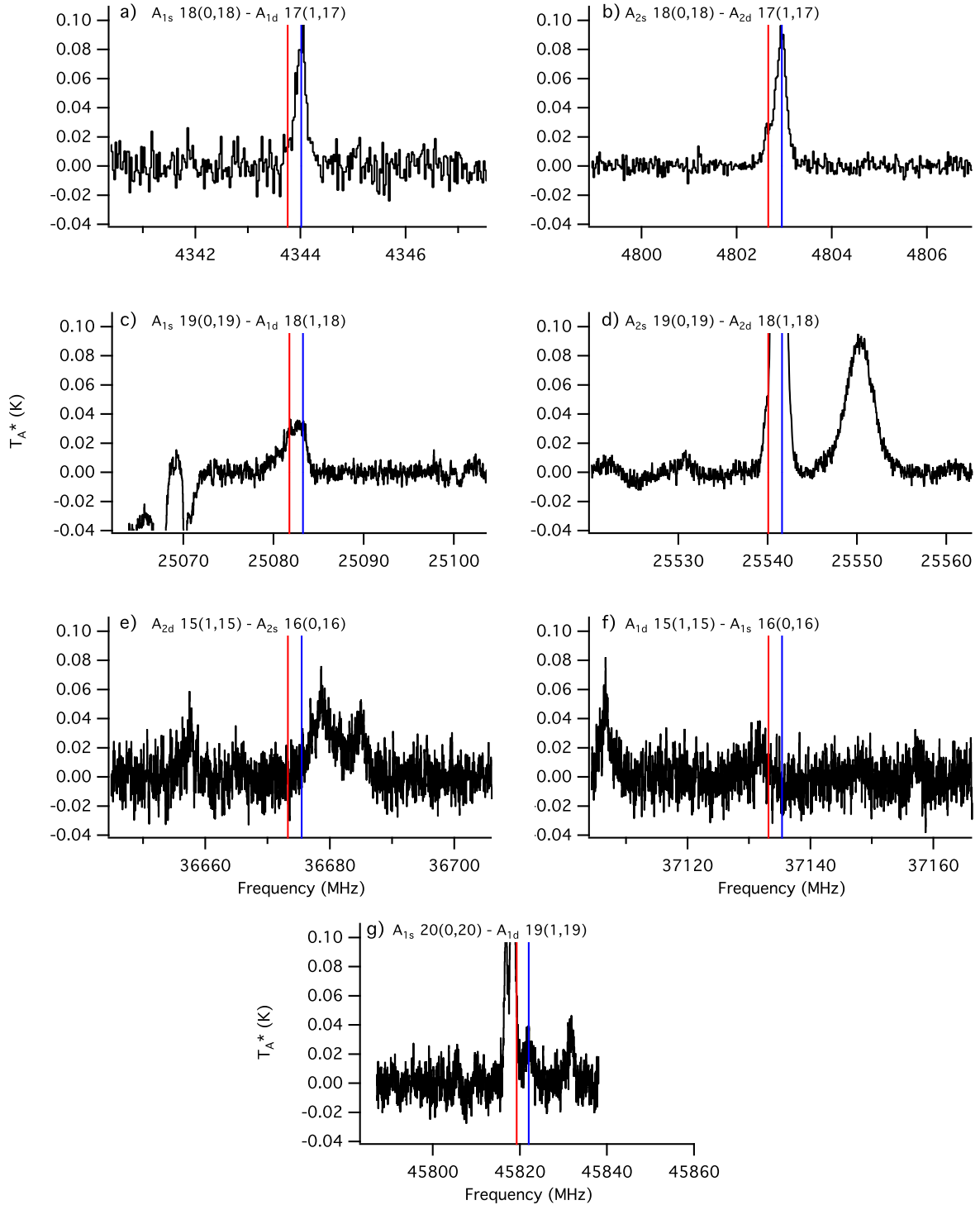}
\label{detects}
\end{figure}

Targeted transitions of HNCNH are shown in Table \ref{transitions} which lists each transition's  rotation-torsion doublet transition quantum numbers, calculated rest frequency (MHz), logarithm of the Einstein A coefficient, lower state energy (cm$^{-1}$), and Gaussian fitted astronomically measured line intensity (mK) and FWHM line widths (km s$^{-1}$) in the observed spectra.  The table provides Gaussian fit parameters for peak intensity and line width for components at v$_{LSR}$ = +64 km s$^{-1}$ and +82 km s$^{-1}$.  Molecular material at these two velocities are characteristic of clouds that lie within the GBT beam along the same line of sight (see e.g. Remijan et al.\ 2008b and references therein).  Figure 2 shows the spectra for the observed passbands.  The red and blue vertical lines indicate the location of the emission feature at LSR velocities of +64 km s$^{-1}$ and +82 km s$^{-1}$, respectively.  We find four unblended lines of HNCNH in this region (panels a) - c) and g)). One frequency range was unobservable by the GBT (16 GHz) while two spectral feature were contaminated by transitions of CH$_3$OH (panels d) \& g)). Finally, one transition was not observed due to lack of frequency coverage (46.2 GHz).  Of the observable transitions, we show clear detections of both the +64 km s$^{-1}$ and +82 km s$^{-1}$ components at 4.3, 4.8 and 25.8 GHz, and the +64 km s$^{-1}$ component at 45.8 GHz (Figure \ref{detects}).  The four line detections and two non-detections are consistent with only masing lines being detectable, as discussed in $\S$3.

\begin{deluxetable}{c r c c c c c c}
\tablewidth{0pt}
\tablecaption{HNCNH Rotation-Torsion Doublets, Spectroscopic, and Observed Astronomical Parameters}
\tablecolumns{9}
\tablehead{
& & & &  \multicolumn{2}{c}{64 km s$^{-1}$} & \multicolumn{2}{c}{82 km s$^{-1}$} \\
\cline{5-8} \vspace{-1em} \\
\colhead{Transition} & \colhead{Frequency\tablenotemark{a}} & \colhead{log$_{10}$($A_{ij}$)} & \colhead{E$_u$} &  Intensity\tablenotemark{b} & FWHM\tablenotemark{b} &  Intensity\tablenotemark{b} & FWHM\tablenotemark{b}  \\
\colhead{$\Gamma ^{\prime}$ J$^{\prime}$(K$^{\prime} _a$,K$^{\prime} _c$) - $\Gamma ^{\prime \prime}$ J$^{\prime \prime}$(K$^{\prime \prime} _a$,K$^{\prime \prime} _c$)} & \colhead{(MHz)} & \colhead{} & \colhead{(cm$^{-1}$)} & \colhead{(mK)} & \colhead{(km s$^{-1}$)} & \colhead{(mK)} & \colhead{(km s$^{-1}$)}} 
\startdata
$A_{1s}$ 18(0,18) - $A_{1d}$ 17(1,17) &	4344.017	  &     -9.10 & 118.2475 & 85(5) & 16(1)  & 13(6) & 8(5)\\
$A_{1d}$ 16(1,16) - $A_{1s}$ 17(0,17) &	16395.527 &	-7.35 & 106.3484 &  \nodata \tablenotemark{c} & \nodata & \nodata \tablenotemark{c} & \nodata \\
$A_{1s}$ 19(0,19) - $A_{1d}$ 18(1,18) &	25083.268 &	-6.81 & 131.3845 & 27(4) &  14.0(0.5) & 28(1) & 22(2)\\
$A_{1d}$ 15(1,15) - $A_{1s}$ 16(0,16) &	37135.396 &	-6.28 & 95.2854   &  $\le12$ & \nodata & $\le12$ & \nodata \\
$A_{1s}$ 20(0,20) - $A_{1d}$ 19(1,19) &	45822.046 &	-6.03 & 145.2126 & 25(1) &  13(2) & \nodata \tablenotemark{d} & \nodata\\
					    		     &		 	  &		  &		        &	 	    \\			
$A_{2s}$ 18(0,18) - $A_{2d}$ 17(1,17) & 4802.952	  &     -8.97 & 118.2548 & 89(2) & 15.9(0.4)   & 23(2) & 12(1) \\
$A_{2d}$ 16(1,16) - $A_{2s}$ 17(0,17) & 15936.094 &     -7.38 & 106.3405 & \nodata \tablenotemark{c} & \nodata & \nodata \tablenotemark{c} & \nodata \\
$A_{2s}$ 19(0,19) - $A_{2d}$ 18(1,18) & 25541.620 &     -6.79 & 131.3918 & \nodata \tablenotemark{d} & \nodata  & (50)\tablenotemark{f} &  (12)\tablenotemark{f}\\
$A_{2d}$ 15(1,15) - $A_{2s}$ 16(0,16) & 36675.469  &	-6.30 & 95.2776   & $\le11$ & \nodata & $\le11$ & \nodata \\
$A_{2s}$ 20(0,20) - $A_{2d}$ 19(1,19) & 46279.903 &     -6.01 & 145.2199 & \nodata \tablenotemark{c} & \nodata & \nodata \tablenotemark{c} & \nodata \\
\enddata
\tablenotetext{a}{All lines except those at 4 GHz have been experimentally measured with (Obs - Calc) frequency uncertainties $\le30$ kHz.  Uncertainties on the 4 GHz lines are calculated to be $\le30$ kHz, type A, k = 2 (2$\sigma$) \citep{Taylor1994}.}
\tablenotetext{b}{The uncertainties for the intensities and linewidths are type B, k = 1 (1$\sigma$) \citep{Taylor1994}.}
\tablenotetext{c}{No data}
\tablenotetext{d}{Transition contaminated by CH$_3$OH emission.}
\tablenotetext{f}{Estimated based on a gaussian fit to the low-frequency shoulder of the CH$_3$OH emission at +82 km s$^{-1}$}
\label{transitions}
\end{deluxetable}

\begin{figure}
\caption{Energy level structure of the relevant transitions for the 4.8 GHz maser (top) and 36.6 GHz line (bottom).  Energy levels are ordered by increasing energy, but are not drawn to scale.  Allowed transitions are indicated by arrows.}
\plotone{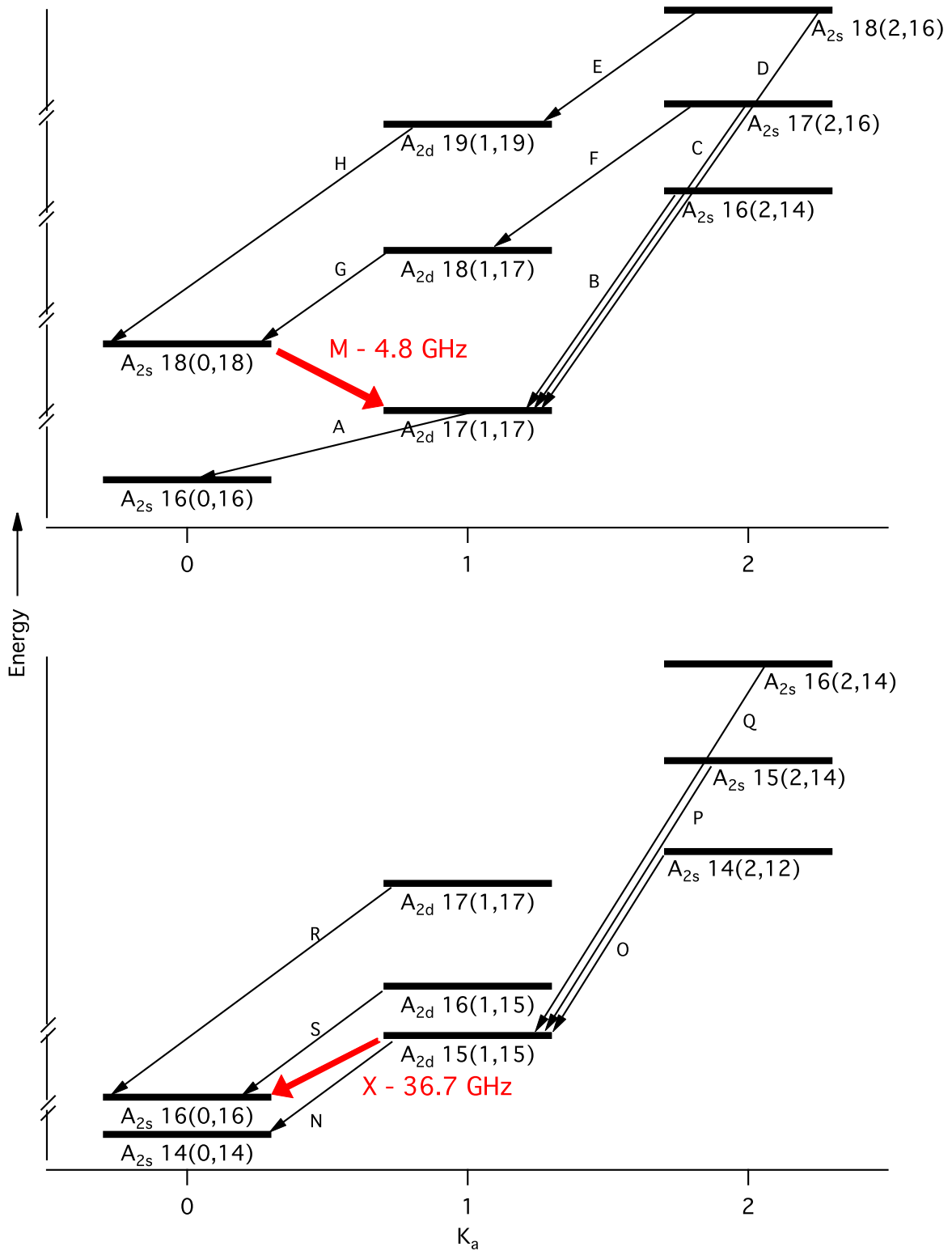}
\label{energylevels}
\end{figure}

\begin{deluxetable}{c r c c r r}
\tablewidth{0pt}
\tablecaption{Spectroscopic Line Parameters for the Transitions Labeled in Figure \ref{energylevels}}
\tablecolumns{6}
\tablehead{
\colhead{} & \colhead{Frequency} & \colhead{Transition} &  \colhead{log$_{10}$($A_{ij}$)} & \colhead{E$_{l}$} & \colhead{E$_{u}$}  \\
\colhead{Label}           & \colhead{(MHz)}             & \colhead{$\Gamma ^{\prime}$ J$^{\prime}$(K$^{\prime} _a$,K$^{\prime} _c$) - $\Gamma ^{\prime \prime}$ J$^{\prime \prime}$(K$^{\prime \prime} _a$,K$^{\prime \prime} _c$)} & \colhead{} & \colhead{(cm$^{-1}$)} &  \colhead{(cm$^{-1}$)}} 
\startdata
A & 720713.219    & $A_{2d}$ 17(1,17) - $A_{2s}$ 16(0,16) & -2.39 & 94.054    & 118.094\\
B & 751698.489    & $A_{2s}$ 16(2,14) - $A_{2s}$ 17(1,17) & -2.72 & 118.094  &  143.167\\
C & 1104054.302 & $A_{2s}$ 17(2,16) - $A_{2s}$ 17(1,17) & -1.85 & 118.094  &  154.921\\
D & 1477122.607 & $A_{2s}$ 18(2,16) - $A_{2s}$ 17(1,17) & -1.74 & 118.094  &  167.365\\
E & 710204.729    & $A_{2s}$ 18(2,16) - $A_{2d}$ 19(1,19) & -2.79 & 143.676 &  167.362\\
F & 730813.907    & $A_{2s}$ 17(2,16) - $A_{2d}$ 18(1,17) & -2.75 & 130.544 &  154.921\\
G & 368437.597   & $A_{2d}$ 18(1,17) - $A_{2s}$ 18(0,18) & -2.98 & 118.254  &  130.544\\
H & 762114.922   & $A_{2d}$ 19(1,19) - $A_{2s}$ 18(0,18) & -2.32 & 118.254  &  143.676\\
{\color{red} \textbf{M}} & {\color{red} \textbf{4802.956}}       & {\color{red} \textbf{$A_{2s}$ 18(0,18) - $A_{2d}$ 17(1,17)}} & {\color{red} \textbf{-8.97}} & {\color{red} \textbf{118.094}} & {\color{red} \textbf{118.254}}  \\
& \\
N & 679302.047   & $A_{2d}$ 15(1,15) - $A_{2s}$ 14(0,14) & -2.47 & 72.618  &  95.277\\
O & 793191.061   & $A_{2s}$ 14(2,12) - $A_{2d}$ 15(1,15) & -2.66 & 95.277  &  121.735\\
P & 1104104.631 & $A_{2s}$ 15(2,14) - $A_{2d}$ 15(1,15) & -1.85 & 95.277  &  132.106\\
Q & 1435736.275 & $A_{2s}$ 16(2,14) - $A_{2d}$ 15(1,15) & -1.74 & 95.277  &  143.168\\
R & 720713.219   & $A_{2d}$ 17(1,17) - $A_{2s}$ 16(0,16) & -2.39 & 94.054  &  118.094\\
S & 368445.962   & $A_{2d}$ 16(1,15) - $A_{2s}$ 16(0,16) & -2.98 & 94.054  &  106.343\\
{\color{red}\textbf{X}} & {\color{red} \textbf{36675.469}}     & {\color{red} \textbf{$A_{2d}$ 15(1,15) - $A_{2s}$ 16(0,16)}} & {\color{red} \textbf{-6.30}} & {\color{red} \textbf{94.054}} & {\color{red} \textbf{95.277}} \\
\enddata
\label{energylabels}
\end{deluxetable}

\clearpage

\section{Discussion}

The initial identification of HNCNH was based on only the two lines at 4.3 and 4.8 GHz. In order to quantify the probability of coincidental overlap of features at these frequencies which could lead to a possible mis-assignment, we used the method outlined in Neill et al. (2012).  We find 37 observed transitions within a representative window of 200 MHz of PRIMOS at C-band (4 - 6 GHz), 14 of which are within $\pm$50\% of the intensity of the detected features.  This line density is typical of the PRIMOS survey in this frequency range.  Of these 14 features, 4 were in emission.  If we then assume a conservative FWHM line width of 25 km s$^{-1}$ (almost twice our measured FWHM) we calculate the probability of a single line falling coincidentally within one FWHM of our line to be 0.75\%.  For two detected lines, that probability drops to 0.002\%.  We find this is both compelling evidence for a clear detection and an excellent example of the power of the GBT in new molecule detections at centimeter wavelengths, where line confusion is drastically reduced relative to the millimeter and sub-millimeter regimes.  

Initially, the large beamwidths and high-energy nature of the observed lines pointed to emission arising from the hot, extended source that surrounds the cold LMH and couples well to the beam at these frequencies.  An LTE analysis, however, showed excessively high column densities for thermal emission arising from these lines.  In order to further verify the identification, and to understand the unusually high intensities of the lines, we examined higher frequency transitions.  Unfortunately, the GBT has no receiver that operates between 15-16 GHz; we detected an emission feature at 25.0 GHz, but its doublet transition at 25.5 GHz is blended with the 9(2,7)-9(1,8) transition of CH$_3$OH at 25.541 GHz; we detected no emission features at 36.6 and 37.1 GHz; and a feature was identified at 45.8 GHz, the highest frequency line currently covered in the PRIMOS survey.  The lack of observed transitions at 36.6 and 37.1 GHz was initially troubling as there has been a firmly established series of requirements for an interstellar detection, including that no transition which should be of an observable intensity be ``missing" (Snyder et al.\ 2005).  However, an examination of the rotational energy level structure of HNCNH offers an explanation.

Figure \ref{energylevels} shows energy level diagrams containing transitions that are relevant to the observed masing lines at 4.8 GHz (labeled ``M", Figure 2(a)) and 36.7 GHz (labeled ``X", Figure 2(b)).  The quantum mechanical parameters associated with transitions in Figure 2 are given in Table \ref{energylabels}.  In the case of the 4.8 GHz line ``M", the upper energy level ($A_{2s}$ 18(0,18)) is coupled to higher energy levels via fast millimeter and sub-millimeter transitions; the only downward transition out of this level is very slow, with an Einstein A coefficient of $\sim10^{-9}$ s$^{-1}$.  Downward and upward transitions out of the lower level ($A_{2d}$ 17(1,17)), labeled ``A - D", however, occur rapidly with Einstein A values between 0.2 - 2 x 10$^{-2}$  s$^{-1}$.  This creates a population inversion between the $A_{2s}$ 18(0,18) and $A_{2d}$ 17(1,17) levels, leading to the observed masing at 4.8 GHz.  An analogous process occurs for the torsion doublet, causing masing at 4.3 GHz.  A similar energy structure exists for the 25.5 GHz and 46.7 GHz transitions and their torsion doublets, leading to masing in these transitions as well.

For the 36.7 GHz line (labelled ``X", Figure 2(b)), the upper state ($A_{2d}$ 15(1,15)) is again coupled to higher energy levels via fast transitions, with the 36.7 GHz transition being the slow emission path to $A_{2s}$ 16(0,16).  In this case, however, a fast downward transition is allowed to the $A_{2s}$ 14(0,14) state (transition labelled ``N").  Population inversion is not achieved because of the presence of this drain, and the 36.7 GHz transition does not mase. A similar scenario holds for the 15.9 GHz transitions, as well as their torsion doublets.  In summary, an analysis of the energy levels of HNCNH predicts a population inversion resulting in masing for the torsion doublets at 4, 25, and 46 GHz, with no masing by the doublets at 15 and 36 GHz.  Furthermore, the 4 GHz doublets should show a greater degree of inversion due to the longer lifetime of the transitions (Table \ref{transitions}) than the 25 or 46 GHz transitions, resulting in more intense lines, which is in excellent agreement with our observations.  

The observed maser emission is relatively weak and displays no line narrowing, suggesting that these are unsaturated maser lines.  As collisional and radiative pumping rates for HNCNH are unknown, a full pumping model was not possible and is beyond the scope of this work. A zeroth-order calculation of the critical density for the fast transitions connecting the maser transitions was, however, performed.  Using collisional coefficients for similar transitions in HNCO, we find critical densities required to thermalize these transitions on the order of 10$^8$ - 10$^{10}$ cm$^{-3}$, far higher than those typically associated with the region.  Given this, and that the Sgr B2(N) region is known to contain a broad range of favorable radiative excitation conditions, a radiative excitation mechanism for these transitions is possible.  The current observations, however, do not provide a definitive answer to the excitation question, and further investigation is warranted. A mapping study of the 4 GHz lines in Sgr B2(N) would be extremely beneficial to elucidating the environment these signals are arising from, including their spatial correlation with other known maser such as CH$_3$OH, and therefore the most likely excitation mechanisms.  

%As collisional and radiative pumping rates for HNCNH are unknown, a full pumping model was not possible, however, the observed maser emission is relatively weak and displays no line narrowing, suggesting that these are unsaturated maser lines.  The temperatures in this source are great enough to populate the ground state of the maser lines, which can then be pumped upwards by far-IR or collisional excitation.  The hot, extended region surrounding the LMH could also contribute to the excitation. A mapping study of the 4 GHz lines in Sgr B2(N) would be extremely beneficial to elucidating the environment these signals are arising from and therefore the most likely excitation mechanisms.

We have also carried out an analysis of the expected intensity of HNCNH lines in this frequency region based on a thermal population of HNCNH.  Assuming local thermodynamic equilibrium (LTE) conditions, we calculate a column density of HNCNH giving rise to the 4 GHz maser lines of $\sim$2 x 10$^{16}$ cm$^{-2}$ at 150 K, although this value is largely invariant over temperatures from 80 - 500 K.  Given this column density, the expected intensity of the 36.7 GHz transition would be $\sim$740 mK.  Further, at a column density of $\sim$2 x 10$^{13}$ cm$^{-2}$ ($\sim$10\% of NH$_2$CN in this source \citep{Nummelin2000}), we would expect the most intense lines in this region to be less than 1 mK for rotational temperatures above 80 K.  Given this, and the  lack of any detection at 36.7 GHz with an RMS noise level of $\sim$10 mK, we conclude these transitions are not arising from a thermal population.

  This is consistent with our non-detection of transitions which are not masing.  In gas above 80 K, the most intense lines of HNCNH at LTE would fall at millimeter and sub-millimeter wavelengths.  Numerous high-resolution, high-sensitivity spectral line surveys of this region have been carried out (see Wang et al. 2011, Tercero et al. 2010 \& Nummelin et al. 2000 and refs. therein) covering the these transitions up to $\sim$1.8 THz.  With an abundance of 10\% NH$_2$CN, the strongest transitions under LTE emission at 80 K would be on the order of the RMS noise level of the survey observations.  With increasing temperature, the expected line strengths decrease rapidly.  Thus, at these column densities, a thermal population of HNCNH would likely not be detectable in even the most sensitive line surveys to date.

%We have also carried out an analysis of the expected intensity of HNCNH lines in this frequency region based on a thermal population of HNCNH at a column density of $\sim$ 2 x 10$^{13}$ cm$^{-2}$ ($\sim$ 10\% of NH$_2$CN in this source \citep{Nummelin2000}).  We expect the most intense lines in this region to be less than 1 mK for rotational temperatures above 80 K.  This is consistent with our non-detection of transitions which are not masing.  In gas above 80 K, the most intense lines of HNCNH in local thermodynamic equilibrium (LTE) would fall at millimeter and sub-millimeter wavelengths.  Numerous high-resolution, high-sensitivity spectral line surveys of this region have been carried out (see Wang et al. 2011, Tercero et al. 2010 \& Nummelin et al. 2000 and refs. therein) covering the these transitions up to $\sim$1.8 THz.  With an abundance of 10\% NH$_2$CN, the strongest transitions under LTE emission at 80 K would be on the order of the RMS noise level of the survey observations.  With increasing temperature, the expected line strengths decrease rapidly.  Thus, at these column densities, a thermal population of HNCNH would likely not be detectable in even the most sensitive line surveys to date.

This is in itself remarkable.  A heroic effort has been performed in recent years on compiling molecular inventories and facilitating new molecule detections using state-of-the-art millimeter and sub-millimeter observatories and molecular line surveys.   These surveys have provided invaluable information on the physical and chemical conditions present in the ISM and their high scientific value is unquestionable.  They suffer, however, from extremely high line densities and long integration times, relative to lower-frequency telescopes.  Because of this, identifications of new molecular species, especially those in low abundance, can be difficult due to the line confusion at the noise floor and the degree of coincidental overlap of target lines with other molecular transitions.  

In this work, we have shown that a molecule which would be undetectable via transitions arising from a thermal population can be identified via maser transitions at centimeter wavelengths.  This detection was possible only through low frequency surveys of a chemically rich region by the GBT through a dedicated project (PRIMOS). The number of transitions which display significant maser activity in any given molecule is likely to be small, but weak masing behavior has been observed in a number of molecules such as H$_2$CO and NH$_3$ \citep{Forster1980,Gaume1991}.  In addition, at centimeter wavelengths the lack of line confusion makes definitive identification of a species possible with only a small number of observed transitions.  This may represent a new strategy for searches for key molecules of interest to the astrochemistry and astrobiology communities which have not yet been detected due to their low abundances.

In summary, we have detected four transitions of carbodiimide (HNCNH) towards Sgr B2(N) with very high confidence.  All four signals have been found to be the result of maser activity.  We also report two transitions that were not detected, consistent with HNCNH only being detectable towards Sgr B2(N) through maser lines.  This detection presents a new methodology for searches for interstellar molecular candidates which may be too low in abundance to be detected in thermal emission by modern radio observatories.

\acknowledgments

All authors dedicate this work to Lewis E. Snyder who pioneered the astronomical detection of molecules at radio frequencies.  B.A.M. gratefully acknowledges funding by an NSF Graduate Research Fellowship, and M. Emprechtinger for helpful discussions regarding masers.  We thank the anonymous referee for very helpful comments.  The National Radio Astronomy Observatory is a facility of the National Science Foundation operated under cooperative agreement by Associated Universities, Inc.

\end{document}